\documentclass[twocolumn,10pt]{asme2ej}
\usepackage{verbatim}
\usepackage{graphicx} %% for loading jpg figures
\usepackage{amsmath}

\usepackage{color,soul}

\title{Fisher Identifiability Analysis of Longitudinal Vehicle Dynamics}

\author{Aaron Kandel\thanks{Address all correspondence to this author.} 
    \affiliation{
	PhD Student\\
	Department of Mechanical Engineering\\
	University of California\\
	Berkeley, California 94704\\
    Email: aaronkandel@berkeley.edu
    }	
}

\author{Mohamed Wahba
    \affiliation{
    PhD Student\\
    Department of Mechanical Engineering\\
	The Pennsylvania State University\\
	University Park, PA 16801\\
        Email: wahba15@gmail.com
    }
}

\author{Hosam K. Fathy  
    \affiliation{
    Professor\\
        Department of Mechanical Engineering\\
        University of Maryland\\
        College Park, MD 20742\\
        Email: hfathy@umd.edu
    }
}

\begin{document}

\maketitle    

%%%%%%%%%%%%%%%%%%%%%%%%%%%%%%%%%%%%%%%%%%%%%%%%%%%%%%%%%%%%%%%%%%%%%%
\begin{abstract}
{\it This paper investigates the theoretical Cram\'er-Rao bounds on estimation accuracy of longitudinal vehicle dynamics parameters.  This analysis is motivated by the value of parameter estimation in various applications, including chassis model validation and active safety.  Relevant literature addresses this demand through algorithms capable of estimating chassis parameters for diverse conditions.  While the implementation of such algorithms has been studied, the question of fundamental limits on their accuracy remains largely unexplored. We address this question by presenting two contributions.  First, this paper presents theoretical findings which reveal the prevailing effects underpinning vehicle chassis parameter identifiability. We then validate these findings with data from on-road experiments.  Our results demonstrate, among a variety of effects, the strong relevance of road grade variability in determining parameter identifiability from a drive cycle.  These findings can motivate improved experimental designs in the future.
}
% This paper investigates the theoretical bounds on estimation accuracy of longitudinal vehicle dynamics parameters.  This analysis is motivated by the value of parameter estimation in various applications, including chassis model validation and active safety.  Relevant literature addresses this demand through algorithms capable of estimating chassis parameters for diverse conditions.  While the implementation of such algorithms has been studied, the question of fundamental limits on their accuracy remains largely unexplored.  We address this question by presenting two contributions.  First, this paper presents theoretical findings which reveal the prevailing effects underpinning vehicle chassis parameter identifiability.  Principal among these effects is the strong relevance of road grade variability in determining parameter identifiability from a drive cycle.  Finally, we validate these insights with on-road experiments.
\end{abstract}

%%%%%%%%%%%%%%%%%%%%%%%%%%%%%%%%%%%%%%%%%%%%%%%%%%%%%%%%%%%%%%%%%%%%%%
\section{Introduction}

This paper explores the impact of drive cycle characteristics on the identifiability of longitudinal
vehicle chassis parameters. From a conventional engineering perspective, vehicle
chassis parameter estimation is an important exercise for identifying and controlling relevant vehicle models. For example, online mass
estimation has significant importance to active vehicle safety systems \cite{Papelis00}.
Similarly, drag estimation can provide utility in quantifying the
benefits associated with a heavy-duty vehicle’s participation in a vehicle platoon \cite{platoon00}.
Furthermore, as connected and automated vehicle systems are studied in greater detail,
online parameter estimation algorithms will become increasingly important, as the
function of connected and automated vehicle (CAV) system optimization will likely depend on accurate knowledge of
parameters including vehicle mass, drag, and rolling resistance coefficients. This
consideration becomes even more important when considering vehicles with highly
variable loads, including freight trucks \cite{heavdut00, heavdut01}.

The literature presents numerous vehicle parameter estimation algorithms. For example, research by Bae et al. \cite{Bae00} applies a recursive least-squares algorithm (RLS) to obtain online estimates of
longitudinal parameters including vehicle mass from experimental data. Work by
Fathy et al. explores online mass estimation using a supervisory algorithm which identifies predominantly
longitudinal vehicle motion \cite{Fathy00}. Vahidi et al. \cite{Vahidi00} estimate both vehicle mass and road grade through RLS utilizing multiple forgetting factors. Rhode et al.  \cite{Rhode00} 
evaluate a generalized version of total RLS based on its experimental performance
in estimating longitudinal parameters including rolling resistance coefficients and vehicle
mass. The authors of this paper also
explore offline nonlinear least-squares longitudinal chassis parameter estimation within
the context of quantifying the impact of terrain variability on the identifiability of such
parameters \cite{Kandel00}.

Vehicle chassis parameter estimation not limited to the
use of longitudinal vehicle dynamics. For example, Rajamani et al. \cite{Rajamani00} show that
suspension dynamics can be utilized to effectively estimate vehicle chassis parameters
including vehicle mass. Pence, Fathy, and Stein  \cite{Pence00} demonstrate that vehicle mass can be estimated inexpensively from base excitation suspension dynamics. Reina et al.  \cite{Reina00} utilize lateral vehicle dynamics with
model-based estimation to estimate vehicle states and
mass.  

Exploring factors which affect parameter identifiability dates back to a landmark paper by Bellman and Astrom \cite{Bellman00}.
Examining limitations related to the identifiability of model parameters provides an
opportunity to observe new insights related to the nature of such estimation problems.
Such insights can address fundamental questions like ``What conditions of an
on-road experiment impose the biggest limitations on chassis parameter estimation
accuracy?", or ``In what ways does the evolution of the drive cycle affect the identifiability of vehicle chassis parameters?" We can also attempt to quantify the impact of the effects
which are present in the scenarios posed by such questions, including for example
quantifying the impact of terrain variability on vehicle chassis parameter identifiability. This is where our paper’s motivation lies.

The literature recognizes the
impact of the design of a vehicle experiment on the accuracy of the resulting chassis
parameter estimates. The SAE testing standard for
longitudinal vehicle chassis parameter estimation outlines several basic experimental
conditions which facilitate estimation accuracy \cite{SAE00}. Within another context, work by Muller et al. \cite{muller00} demonstrates that the addition of speed bumps to a road test allows
relevant suspension parameters to be estimated with greater precision. While such work recognizes effects which facilitate estimation accuracy, there is a need to provide theoretical justification for such observations.  Pursuing such theoretical justification could potentially reveal new insights which can improve experimental design.  For instance, an understanding of the impact of terrain variability on longitudinal chassis parameter identifiability is fairly unexplored.  %this area\cite{Kandel00} by addressing these open questions in greater detail.

The Fisher information metric is an effective means for quantifying parameter identifiability \cite{Norton00}.
The invertibility of the Fisher information matrix
indicates whether the experiment in question can yield uniquely identifiable
parameter estimates. Specifically, the Cram\'er-Rao theorem states that the best 
parameter estimation covariance achievable by an unbiased estimator is given by the
inverse of the matrix-formulation of Fisher information. In this paper, we present a theoretical analysis which reveals the underlying effects and limitations on the best achievable estimation accuracy for longitudinal vehicle dynamics. For an appropriate estimation algorithm, our analysis leverages Fisher Information to mathematically show the form of the Cramer-Rao lower bounds on longitudinal parameter estimates. We use these expressions to show fundamental insights which can motivate improved experiment design.
%The preliminary sections of
%this paper are focused on deriving analytic expressions for the Fisher information matrix
%for a longitudinal vehicle dynamics model with respect to vehicle mass, drag
%coefficient, and rolling resistance parameters.  From these analytic derivations, we develop insights into the factors affecting chassis parameter identifiability.   

%After validating these results against those obtained from conventional numeric approximations of Fisher information, we conduct several Monte Carlo simulation studies based on varying driving conditions. These simulations validate insights from our theoretical analyses. 

The remaining organization of this paper is as follows. Section II details our
formulation of the longitudinal vehicle dynamics. Section III describes the format of
the Fisher information metric corresponding to the longitudinal dynamics.  In Section IV, we derive an analytic Fisher matrix for a linear estimation case.
We describe our methods for obtaining analytic expressions for Fisher information with nonlinear least-squares in
Section V, including validation of our derivations through simulation and on-road experiments. Finally, after section V our conclusion summarizes and
synthesizes insights from the paper's findings. 

\section{Longitudinal Vehicle Dynamics Model}

The foundation of this paper’s analysis utilizes a nonlinear longitudinal vehicle
dynamics model adopted from the literature \cite{Rajamani01}. This model formulation allows us to
evaluate the identifiability of vehicle mass, drag coefficient, and rolling resistance
coefficient parameters.
The nonlinear state-space model is given by (\ref{eqn::longdyn1}-\ref{eqn::longdyn2}).

\begin{align}
\dot{x} &= v \label{eqn::longdyn1} \\
\dot{v} &= \frac{1}{m}[F - \frac{1}{2}\rho C_d A_{ref}v^2] - \mu g \cos(\beta) - g \sin(\beta)\label{eqn::longdyn2}
\end{align}
where $x$ is the vehicle's position, $v$ is the vehicle velocity, $\dot{v}$ is the vehicle acceleration, $F$ is the propulsion force, and $\beta$ is the road grade.   Table 1 describes other relevant parameters of this model. 

\begin{table}
\caption{Longitudinal Dynamics Model}
\begin{center}
\label{table_ASME}
\begin{tabular}{lcc} 
& & \\ % put some space after the caption
 %& \multicolumn{2}{l}{Type} \\ \cmidrule{2-7}
 Value & Description & Units\\ 
 \hline
 $m$ & vehicle mass & kg\\
 \hline
 $C_d$ & drag coefficient & - \\
 \hline
 $\mu$  & rolling resistance coefficient & - \\ 
 \hline
 $A_{ref}$ & frontal area & $m^2$ \\
 \hline
 $\rho$ & air density & [$\frac{kg}{m^3}$] \\
 \hline
 $g$ & gravitational constant & $\frac{m}{s^2}$ \\
\end{tabular}
\end{center}
\end{table}

\section{Formulation of Fisher Information}
Fisher information is a metric which quantifies how much
information a data set contains about a set of relevant model parameters \cite{Norton00}. In formulating our Fisher analysis, we first define relevant system output and
longitudinal chassis parameters from the model described by (\ref{eqn::longdyn1}-\ref{eqn::longdyn2}). We specifically select
vehicle mass $m$, $C_d$, and $\mu$ to be the chassis parameters which we assume are being
estimated:
\begin{equation}
    \theta = \begin{pmatrix} m \\ C_d \\ \mu \end{pmatrix}
\end{equation}
Fisher information analysis requires selection of a measured output variable.  In this paper, we use longitudinal vehicle velocity as our measured output, as this aligns with our experimental setup. 

The Fisher information matrix quantifies an expected curvature (i.e. Hessian) of the likelihood function specifically around the parameter estimate $\theta$.  In the case of linear least-squares, the Fisher information matrix is computed as:
\begin{equation}
    I(\theta) = \frac{R^TR}{\sigma^2}
\end{equation}
where $R$ is the regressor matrix, and $\sigma^2$ is the sensor variance associated with output measurement noise for an assumed white, independent and identically distributed noise process. This indicates that estimation accuracy depends not only on how we excite the system, but on how well our sensors can accurately measure their intended signals.

For nonlinear least-squares problems, Fisher information can be computed as follows \cite{Thomas00}: 
\begin{equation}\label{itheta}
    I(\theta) = \frac{S^TS}{\sigma^2}
\end{equation}
We note that
\begin{equation}\label{itheta2} % \label{eqn::sens}
    S_{ij} = \frac{\partial y(t_i)}{\partial \theta_j}
\end{equation}
where $i$ indicates the timestep and $j$ indicates the parameter, is the sensitivity of the output with respect to the parameters. This sensitivity can be approximated via finite differences:
\begin{equation}\label{Num:sens}
    \frac{\partial v(t_i)}{\partial \theta_j} = \frac{v(t_i, \theta_j+\epsilon_j) - v(t_i, \theta_j)}{\epsilon_j}
\end{equation}
In section IV, we derive analytic continuous-time expressions for
the output sensitivities given by (\ref{itheta2}). We therefore must approximate the summations in the Fisher information matrix with integrals when using these analytic expressions:
\begin{equation}\label{eqn::CTFI}
    \sum_{i=1}^N S_{i}S_j \rightarrow f \int_{0}^T S_{i}(t)S_{j}(t)dt
\end{equation}
where $T$ is the final time and $f$ is the sampling frequency.

If the Fisher information matrix is positive definite, the model parameters are
said to be uniquely locally identifiable. Furthermore, the Cram\'er-Rao theorem states
that the best achievable estimation covariance is obtained from the inverse of the Fisher information matrix. The diagonal terms of this covariance
matrix are the Cram\'er-Rao lower bounds of the estimation variances for each parameter
estimate in $\theta$.  These estimation error variances quantify the accuracy of the estimator. This paper's simulated and experimental identifiability analyses compare the error bounds obtained from Fisher analyses using both numeric and analytic methods.

\section{Linear Estimation Case}
We perform Fisher analysis for both the full nonlinear vehicle parameter estimation problem and a simplified linear case.

%We begin this paper's technical content by providing a motivating example of the potential impact of terrain variability on parameter identifiability. The need for simplicity in this case is imperative. Even cursory analysis seeking to obtain analytic expressions of the Cramer-Rao estimation error bounds reveals that without proper simplification and assumptions, the complexity of the results rapidly escalates to the point they lose clear interpretation.  
To perform the simplified linear analysis, we start with the basic longitudinal model formulation:
\begin{equation}
    m \dot{v} = F - \frac{1}{2}\rho C_d A_f v^2 - \mu m g \cos(\beta) - m g \sin(\beta) 
\end{equation}
we assume the road grade $\beta$ is sufficiently small such that
\begin{equation}
    m \dot{v} \approx F - \frac{1}{2}\rho C_d A_f v^2 - \mu m g - m g \beta
\end{equation}
Furthermore, we assume rolling resistance $\mu$ is known.  This is a fair assumption: vehicle mass can vary significantly (e.g. heavy-duty vehicles), and aerodynamic drag can likewise change due to inter-vehicle interactions (i.e. platooning), but rolling resistance tends to stay fairly constant. By assuming $\mu$ and $\beta$ are known, we can reformulate the longitudinal model to be conducive to application of ordinary least-squares as follows:
\begin{equation}
    m(\dot{v} + \mu g + g \beta) + \frac{1}{2}\rho C_d A_f v^2 = F
\end{equation}
where we now redefine $x_1(t) = \dot{v} + \mu g + g \beta$, $x_2(t) = v^2$, $y=F$, $\theta_1 = m$, and $\theta_2=\frac{1}{2}\rho C_d A_f$. By using (\ref{eqn::CTFI}) to reformulate the Fisher matrix in continuous time, we obtain the following expression:
\begin{equation}
    I(\theta) \approx \frac{f}{\sigma^2}\begin{bmatrix}\int_{0}^T x_1^2(t)dt & \int_{0}^T x_1(t)x_2(t)dt  \\ \int_{0}^T x_1(t)x_2(t)dt & \int_{0}^T x_2^2(t)dt \end{bmatrix}
\end{equation}
From this point forward, we assume we are considering periodic conditions where $v(t_0) = v(t_f)$ and $\text{elevation}|_{t_0} = \text{elevation}|_{t_f}$. Now, filling in the values for $x_1(t)$ and $x_2(t)$ as:
\begin{equation} % \int v^2 (\dot{v} + \mu g + g \beta) dt =
    \int x_1(t)x_2(t) dt =  \int v^2 \dot{v} dt + \int v^2 \mu g dt + \int v^2 g \beta dt
\end{equation}
Using integration by parts with $u=v^2$ and $dv = \dot{v}dt$, we transform this expression into:
\begin{equation}
    \int v^2 \dot{v} dt  = v^3 |_{t_0}^{t_f} - \int 2 v \dot{v} v dt
\end{equation}
Since we assume periodic conditions, $v^3 |_{t_0}^{t_f} = 0$. This means that
\begin{equation}
    \int v^2 \dot{v} dt  = - \int 2 v \dot{v} v dt = 0.
\end{equation}
Analyzing the next term yields the following progression:
\begin{equation}
    \int v^2 \mu g dt = \mu g \int v^2 dt = \mu g \frac{1}{3}(v^3 |_{t_0}^{t_f}) = 0
\end{equation}
due to the periodicity assumption.

So, these findings indicate that
\begin{equation}\label{eqn::finalXterm}
    \int x_1(t)x_2(t) dt  = g \int v^2 \beta dt
\end{equation}
Finally, we assume the road grade and kinetic energy are sufficiently uncorrelated with each other.  This assumptions renders the off-diagonal terms of the Fisher information matrix (\ref{eqn::finalXterm}) to equal zero. This leaves us with a diagonal Fisher information matrix:
\begin{equation}
    I(\theta) = \frac{f}{\sigma^2}\begin{bmatrix}\int_{0}^T x_1^2(t)dt & 0  \\ 0 & \int_{0}^T x_2^2(t)dt \end{bmatrix}
\end{equation}
Now, we expand the remaining diagonal terms as follows:
\begin{align}
    \int x_1^2(t) dt &= \int(\dot{v} + \mu g + g \beta)^2 dt\\
    &= \int [\dot{v}^2+\mu^2 g^2 + g^2 \beta^2 + 2 \dot{v}\mu g + 2 \dot{v}g \beta + 2 g^2 \mu \beta] dt
\end{align}
Given the periodicity assumption, this simplifies into:
\begin{equation}\label{FI::11}
    \int x_1^2(t) dt =\int \dot{v}^2dt + T \mu^2 g^2 + g^2 \int \beta^2 dt + 2g\int \dot{v}\beta dt
\end{equation}
Finally, we expand the other diagonal term
\begin{equation}\label{FI::22}
    \int x_2^2(t) dt = \int v^4(t) dt
\end{equation}
resulting in the following Fisher information matrix:
\begin{equation}\label{FI::final}
    I(\theta) = \frac{f}{\sigma^2}\begin{bmatrix} \int_0^T(\dot{v}(t)+g\beta(t))^2 dt + T\mu^2g^2& 0  \\ 0 & \int_0^T v^4(t) dt \end{bmatrix}
\end{equation}
This resulting expression allows us to infer several insights about chassis parameter identifiability. 
\begin{itemize}
    \item[\textbullet] First, our ability to estimate the drag coefficient is largely dependent on the kinetic energy of the experiment squared. This supports the existing empirical testing standard \cite{SAE00}, which dictates a vehicle be brought to high velocity and then coasted down slowly.
    
    \item[\textbullet] Vehicle mass identifiability is subject to a host of more varied, interacting effects.  A new insight from  (\ref{FI::final}) is that rolling resistance squared also affects the identifiability of the mass parameter.  Specifically, the larger the rolling resistance coefficient the easier it will be for one to estimate mass from measurements of the vehicle's propulsive force.
    
    \item[\textbullet] Perhaps the most important factor which affects the mass parameter identifiability is the interplay between vehicle acceleration and gravitational acceleration. When expanding this integral out, it is clear the increased levels of road grade variablility (as represented by the integral of $g^2\beta^2$) improve mass identifiability.  However, this is only true insofar as the vehicle's acceleration deviates from pure gravitational acceleration.  In the worst case, vehicle acceleration $\dot{v} = -g \beta(t)$ or is purely dependent on gravity, which can lead the mass estimation error to be considerably large.  This makes intuitive sense. Consider that if one lets the vehicle accelerate and decelerate purely due to gravity up and down a series of hills, the vehicle will accelerate nearly the same way regardless of its mass.  This makes it almost impossible to estimate vehicle mass unless the control input forces the vehicle to accelerate in ways which depart from pure gravitational acceleration. In this case, mean square road grade (the integral of road grade squared) will only improve estimation errors if the vehicle is controlled in such a way as to suppress the dependence of vehicle acceleration on road grade and gravity.% One simple way to frame this would be to have a vehicle drive fast up a hill, then descend slowly down the hill.
\end{itemize}
% Driving backwards?

These insights provide theoretical support to existing empirical testing standards.  In the remainder of this paper, we will explore these effects with more comprehensive theoretical and experimental analyses.  Specifically, we will first derive analytic expressions for the Fisher information matrix when conducting full nonlinear least-squares to estimate vehicle mass, drag coefficient, and rolling resistance coefficient.  From these derivations, we corroborate the results of this exploratory linear analysis. %Finally, after validating our derivations with experiments, we will conduct a series of simulation studies which verify our theoretical findings under diverse driving conditions.

\section{Nonlinear Least-Squares Analysis}
Now, we consider a full nonlinear least-squares problem where we estimate vehicle mass, drag, and rolling resistance coefficient.  Simultaneous estimation of these three parameters is a more complicated process. By deriving analytic expressions for the Fisher information matrix, we can design simulation studies which replicate the conditions we found ideal in the previous section.  In this section, we show that the effects which improve estimation accuracy for the full (i.e. 3 parameter) problem are similar to those we found in the previous section, where we only estimate vehicle mass and drag coefficient. In the following section, we also show experimentalvalidation of our analytic Fisher information matrix which reveals strong correlation between $C_d$ and $-\mu$ sensitivities, which likely explains the agreement between our results in both cases. 

As described in Section III, our format of Fisher information utilizes expressions for the sensitivity of vehicle velocity with respect perturbations in vehicle mass, drag coefficient, and rolling resistance coefficient parameters. Namely, we adopt equations (\ref{itheta}-\ref{itheta2}) exactly. In this section, analytic expressions for the sensitivity of vehicle velocity with respect to perturbation in vehicle mass, $C_d$, and $\mu$ parameters are derived. Then, we assemble the full Fisher information matrix via (\ref{itheta}).

Subsection 4.1 details our derivation of the nominal longitudinal velocity
trajectory, and subsection 4.2 describes our analytic derivations for the sensitivity expressions. Subsection 4.3 demonstrates the accuracy of our analytic expressions through relevant simulation and experimentation.

\subsection{Obtaining the Nominal Velocity Trajectory}
To obtain the nominal velocity trajectory for use in our sensitivity derivations, we start
with the full nonlinear longitudinal dynamics model with generalized, time-varying
propulsion force and road grade:
\begin{multline}
    \dot{v}(t) = \frac{1}{m}[F(t) - \frac{1}{2}\rho C_d A_{ref}(v(t))^2] - \mu g \cos(\beta(t)) \\
    - g \sin(\beta(t))\label{eqn::longdyn3}
\end{multline}
In this paper, we treat propulsion force and road grade as time dependent quantities.  In reality, they both depend on position which itself depends on time.  We approximate them as time varying to simplify our derivations.

Writing (\ref{eqn::longdyn3}) for a drive cycle subject to a flat terrain profile and constant propulsion force yields the following equation:
\begin{equation}
    \dot{v}_0 = \frac{1}{m}[F_0 - \frac{1}{2}\rho C_d A_{ref}v_0^2] - \mu g\label{eqn::longdyn4}
\end{equation}
Here, we take a flat terrain profile as the operating point for a linearization of the model.  If we take the following perturbations:
\begin{align}
    F(t) &= F_0 + \delta F(t) \label{eqn::delF} \\
    \beta(t) &= \beta_0 + \delta \beta(t) \label{eqn::delbet}
\end{align}
with $\beta_0 = 0$ and $F_0$ being the force required for our vehicle to maintain a constant predetermined speed $v_0$ across a flat terrain.  Next, we can subtract (\ref{eqn::longdyn4}) from the expression we get by plugging in (\ref{eqn::delF}) and (\ref{eqn::delbet}) into (\ref{eqn::longdyn3}) to obtain the following equation for $\delta \dot{v}_n = \dot{v}_n - \dot{v}_0$:
\begin{multline}
    \delta \dot{v}_n = \frac{\delta F(t)}{m} - \frac{1}{m}\rho C_d A_{ref} (v_n^2 - v_0^2) - \mu g( \cos(\delta \beta (t))-1) \\
    - g \sin(\delta \beta(t))
\end{multline}
Furthermore, the term $v^2 - v_0^2$ can be linearized into $2v_0\delta v$ when neglecting higher order terms, where $v_0$ is the constant speed obtained by applying the constant propulsion force $F_0$ to the vehicle subject to a flat terrain:
\begin{multline}
    \delta \dot{v}_n = \frac{\delta F(t)}{m} - \frac{2}{m}\rho C_d A_{ref} v_0 \delta v_n - \mu g( \cos(\delta \beta (t))-1) \\
    - g \sin(\delta \beta(t))
\end{multline}
Solving this ordinary differential equation with zero initial conditions for $\delta v_n(t)$ yields the expression for perturbed vehicle velocity relative to steady state, with respect to an arbitrary time-varying terrain profile and propulsion force:
\begin{multline}
    \delta v_n(t) = e^{-\frac{2C_d A_{ref} \rho v_0 t}{m}} \int_0^t \sigma(\tau)\big{[}\frac{\delta F(\tau)}{m} - g(\sin(\delta \beta(\tau)) \\- \mu + \mu \cos(\delta \beta(\tau)))\big{]}d\tau \label{eqn::vn1}
\end{multline}
where
\begin{equation}
    \sigma(t) = e^{\frac{2C_d A_{ref} \rho v_0 t}{m}} \label{eqn::vn2}
\end{equation}
We can add this expression to $v_0$ to obtain the nominal linearized velocity trajectory for
arbitrary time-varying terrain and propulsion force. This overall expression is essential
to the analytic derivation of the sensitivity of velocity with respect to perturbations in
each chassis parameter $m$, $C_d$, and $\mu$.

\subsection{Sensitivity Derivations}
The following subsection details analytic calculations for the sensitivity of velocity
with respect to perturbations in the longitudinal chassis parameter $m$. The same procedure we use for $m$ can be applied to the other chassis parameters, but for the purpose of brevity we omit those derivations.

Our first step is to
rewrite the full longitudinal model with a percent perturbation applied to
the mass parameter. 
\begin{multline}
    \dot{v} = \frac{1}{m(1+\epsilon_m)}\big{[}F(t) - \frac{1}{2}\rho C_d A_{ref} v^2 \big{]} - \mu g \cos(\delta \beta (t)) \\
    - g \sin(\delta \beta(t))
\end{multline}
This added contribution can be effectively linearized to take the following form:
\begin{multline}
    \dot{v} = \frac{1-\epsilon_m}{m}\big{[}F(t) - \frac{1}{2}\rho C_d A_{ref} v^2 \big{]} - \mu g \cos(\delta \beta (t)) \\
    - g \sin(\delta \beta(t)) \label{eqn::sm1}
\end{multline}
The original longitudinal model in (\ref{eqn::longdyn3}) can be subtracted from (\ref{eqn::sm1}) to give the following expression for our new $\delta \dot{v}$:
\begin{multline}
    \delta \dot{v}(t) = -\frac{\epsilon_m}{m}\big{[}(F_0 + \delta F(t)) - \frac{1}{2}C_d \rho A_{ref}(v(t))^2\big{]} \\
    - \frac{C_d}{m}\rho A_{ref} v(t) \delta v(t)
\end{multline}
where
\begin{align}
    v(t) &= v_0 + \delta v_n(t) \label{eqn::vel}\\
    \dot{v}(t) &= \dot{v}_0 + \delta \dot{v}(t) = \delta \dot{v}(t)
\end{align}
and $\delta v_n(t)$ is represented by (\ref{eqn::vn1}-\ref{eqn::vn2}) in Section 4.1. Upon dividing through by $\epsilon_m$, we obtain a simple time varying 1st order ordinary differential equation which we can solve directly for the sensitivity $\frac{\delta v(t)}{\epsilon_m}$:
\begin{multline}
    \frac{\delta \dot{v}(t)}{\epsilon_m} = -\frac{1}{m}\big{[}(F_0 + \delta F(t)) - \frac{1}{2}C_d\rho A_{ref}(v(t))^2\big{]} \\
    - \frac{C_d}{m}\rho A_{ref} v(t) \frac{\delta v}{\epsilon_m}
\end{multline}
where $v(t)$ is sourced from (\ref{eqn::vel}). 

We use the MATLAB symbolic algebra toolbox to analytically solve this ODE for $\frac{\delta v(t)}{\epsilon_m}$.  For the purpose of brevity we omit the final, lengthy, analytic expression for $\frac{\delta v(t)}{\epsilon_m}$.

\subsection{Intuition from Derivations}
We notice the commonalities the sensitivity expressions share with the simple derivation in the previous section.  For instance, upon using a small angle approximation, the term 
\begin{equation}
    m \int_0^T \alpha(t) (\frac{\delta F(t)}{m} - g \beta(t)) dt
\end{equation}
where $\alpha(t)$ includes additional effects, appears several times in the sensitivity expressions (and thus the resulting Fisher information matrix).  This term originates from the perturbed velocity expression we solve for at the beginning of this section.  Considering that $\frac{\delta F(t)}{m}$ is a propulsion induced acceleration term, this shows a commonality between the linear and nonlinear cases in terms of how road grade affects parameter identifiability. In fact, for the mass sensitivity equation, the interaction between propulsion force and gravitational acceleration is squared, similarly to the upper left hand entry of the linear Fisher information matrix (\ref{FI::final}).  This indicates that similar insights prevail in the nonlinear estimation case, namely that the interaction between acceleration and gravitational forces is important in determining parameter identifiability from an experiment.

The remainder of this paper presents simulation studies which validate our sensitivity expressions under various driving conditions.

\subsection{Simulation and Experimental Validation of Sensitivity Expressions}
To validate our analytic sensitivity expressions, we utilize both a simulation study and an experimental analysis.  First, in our simulation study, we compute a representative driving segment with sufficient excitation.  Then, we simulate longitudinal driving along that segment subject to the same longitudinal model parameterization given in \cite{Kandel00} corresponding to a Volvo VNL300 heavy-duty vehicle, namely $m=8875$ kg, $C_d = 0.49$, and $\mu = 0.0056528$. Using the data from this simulated driving, we compute the parameteric sensitivity signals using expressions given in Section 4.2 of this paper and those we obtain through (\ref{itheta2}).  For our analytic expressions, we take the operating points as (1) the average measured velocity and (2) the average measured wheel force.  Figure 1 demonstrates this comparison for 20 minutes of simulated driving.  The sensitivities plotted in Figure 1 are not normalized.  As we can see, even with a relatively large range of velocities the requisite approximations we make in our analytic derivations end up having little effect on the accuracy of our final results.  Furthermore, the Cram\'er-Rao bounds we obtain from this simulation using numerical and analytic sensitivities are in strong agreement. Table 2 lists the estimation error bounds from both numerical and analytic sensitivities.
\begin{table*}[t]
\caption{Cram\'er-Rao Bounds from Simulated and Experimental Validation of Sensitivity}
\begin{center}
\label{table_ASME}
\begin{tabular}{c l l l l}
& & \\ % put some space after the caption
\hline
Estimation Error & Numerical (sim) & Analytic (sim) & Numerical (exp) & Analytic (exp) \\ 
\hline
 Mass $m$ [kg] & 47.30 & 48.00 & 26.91 & 28.89 \\
 Drag Coefficient $C_d$ [-] & 0.0043 & 0.0039 & 0.0108 & 0.0091 \\
 Rolling Resistance $\mu$ [-] & 0.00022 & 0.00020 & 0.00086 & 0.00078\\
\hline
\end{tabular}
\end{center}
\end{table*}
\begin{figure*}
\centering{%
{\includegraphics[trim = 15mm 5mm 15mm 5mm, clip, width=\textwidth]{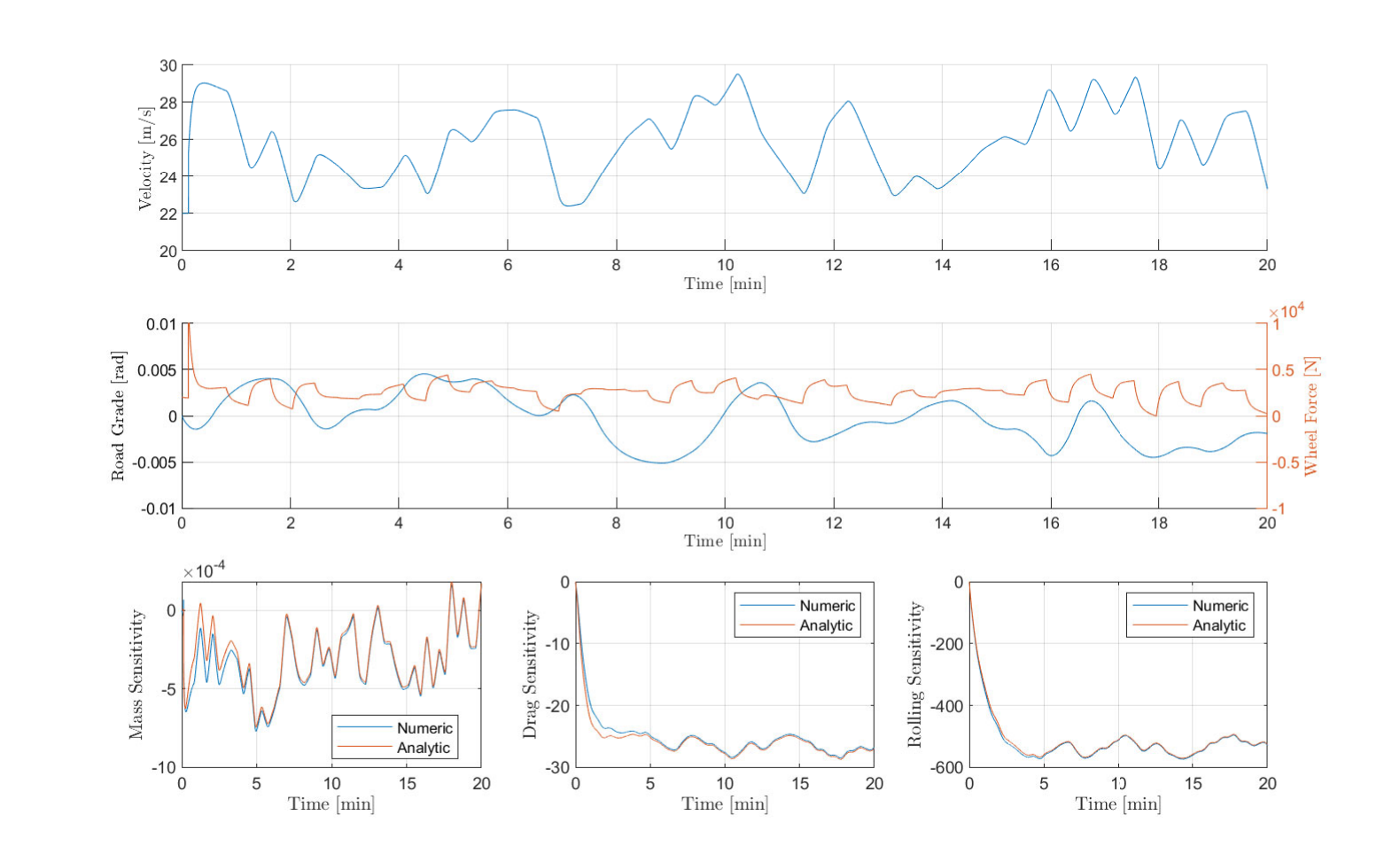}}}\hspace{5pt}
\caption{Simulated validation of analytic sensitivity derivation} \label{sample-figure}
\end{figure*}

For our experimental study, we use data from an instrumented heavy-duty vehicle driving on-road to validate our analytic sensitivity expressions.  The experimental vehicle is instrumented with a final-drive torque sensor, GPS, and a CAN interface which logs data through a Simulink real-time machine. Our use of GPS to measure vehicle velocity negates the requirement for consideration of wheel slip in our dynamical model. The precise details of this experimental setup can be referenced in \cite{Kandel00}.  This setup allows us to record values of wheel torque, road grade, velocity and acceleration, braking indicator, and other useful signals.   To conduct our experimental comparison, we cut the data using the braking indicator to avoid unmodeled braking dynamics.  Then, we compute the sensitivites of the vehicle velocity with respect to each nominal parameter value both numerically (via \ref{Num:sens}) and using our analytic expressions from the previous section.  With both sets of sensitivities, we compute Fisher information using (\ref{itheta}).  Figure 2 demonstrates the results of this comparison.  Overall, we observe strong agreement between numeric and analytic sensitivity signals for highway driving.  Upon computing the Fisher information matrices for both the analytic and numeric cases, our resulting Cram\'er-Rao estimation error lower bounds also agree. This validates the accuracy of our derivations, and shows the necessary approximations have minimal effect on the accuracy of our analytic expressions relative to a more computationally expensive and general approach for computing the sensitivities. Table 2 shows the Cram\'er-Rao bounds obtained from the above data sample. 

In both the simulated and experimental cases, the drag and rolling resistance sensitivities stabilize over time, whereas the mass sensitivity fluctuates throughout the experiment.  This is due to the type of contributions each have directly on the vehicle velocity.  Both drag and rolling resistance present consistently resistive effects on velocity, whereas vehicle mass can increase or decrease the vehicle velocity depending on the interactions between the propulsion force, road grade, and velocity itself.  For instance, consider the case where the vehicle is traveling at constant velocity. If the drag coefficient increases, for a consistent experiment, we would expect the vehicle's velocity to gradually decrease until stabilizing due to the quadratic drag effect. Conversely, if we increase the vehicle mass for a consistent experiment, the vehicle could speed up or slow down depending on whether it is traveling up or downhill.  While the pure gravitational acceleration will be the same with a scale in vehicle mass, the relative impact of the other forces (i.e. drag, rolling resistance, etc..) become smaller, which accentuates the gravitational effect. This can be seen most clearly in the mass sensitivity of the experiment in Figure 2, where the overall velocity does not change a great deal throughout the experiment.  In this case, with consistent downhill driving at the beginning of the experiment the mass sensitivity gradually increases.

\begin{figure*}
\centering{%
{\includegraphics[trim = 15mm 5mm 15mm 5mm, clip, width=\textwidth]{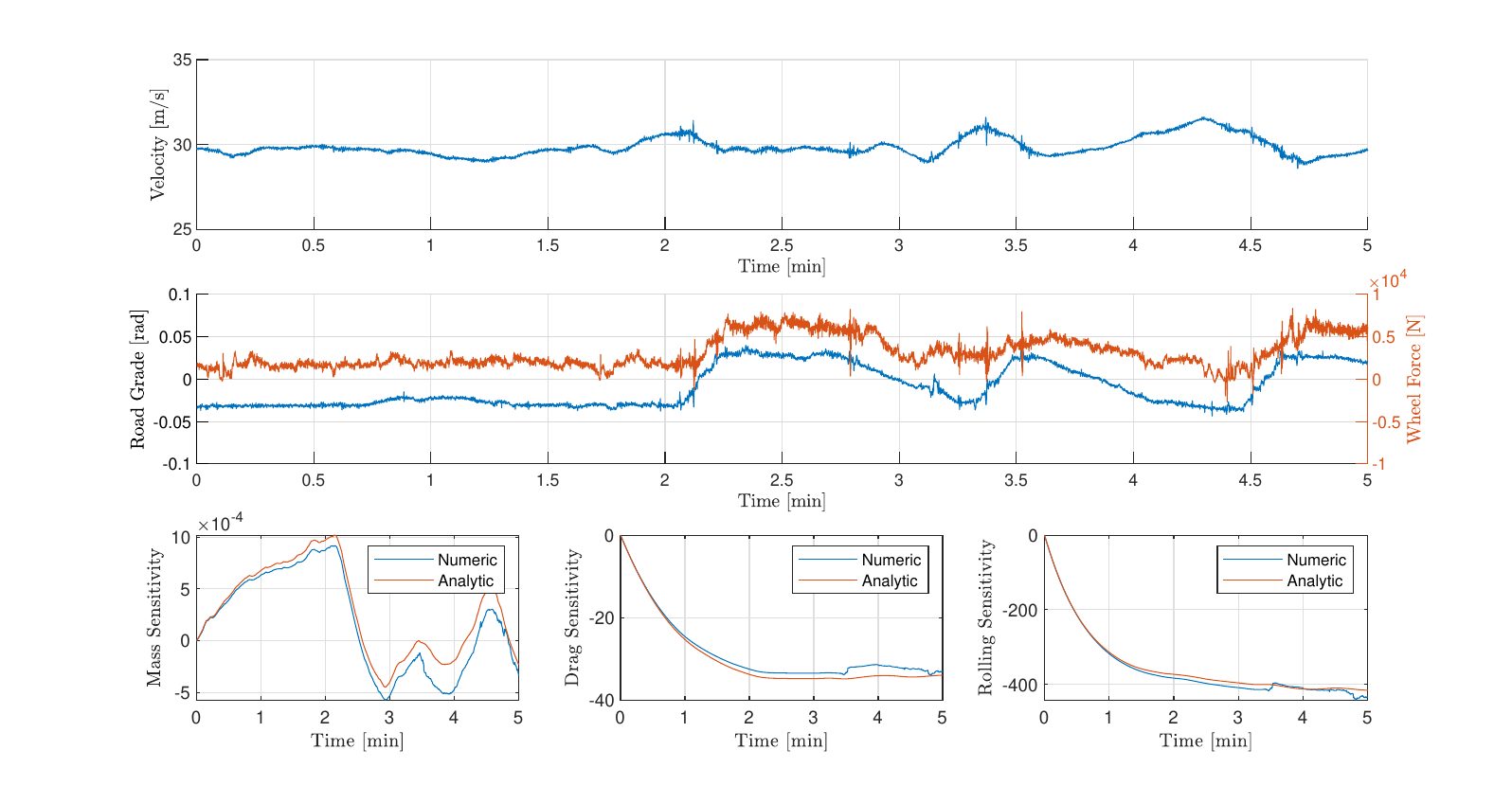}}}\hspace{5pt}
\caption{Experimental validation of analytic sensitivity derivation} \label{sample-figure}
\end{figure*}

\section{Conclusion}
This paper presents two key contributions. These include (i) theoretical insights into the dynamics of chassis parameter identifiability, and (ii) two studies which illustrate the relationships between
terrain variability, velocity, and propulsion force and longitudinal chassis parameter identifiability.

The first application
of these insights pertains directly to the design of on-road experiments for longitudinal
chassis parameter identifiability. Choosing routes for such experiments which possess
the greatest degree of terrain variability could not only improve the accuracy of
parameter estimates obtained from the experiment, but could also enable researchers to
avoid unnecessarily costly instrumentation. Perhaps more significant than
the application to experimental design is the relevance of this paper’s results to
connected and automated vehicle (CAV) systems research. Much of this research relies
on a priori knowledge of relevant vehicle parameters, including vehicle mass, drag, and
rolling resistance coefficients. For actual implementation of CAV system optimization,
these parameters may in fact need to be estimated in real time with online parameter
estimation algorithms, and such algorithms will need to interface with system-wide
optimization. The effectiveness of CAV optimization algorithms will likely depend on
the accuracy of these parameter estimates, and as a result choosing and weighting data 
from segments of road characterized by high terrain variability can improve the
function of such algorithms.

\begin{acknowledgment}
This work is supported by a National Science Foundation Graduate Research Fellowship, and by NSF award \#CMMI-1538300, NSF award \#CMMI-1351146, and ARPA-E award \#DE-AR0000801. 
\end{acknowledgment}

%%%%%%%%%%%%%%%%%%%%%%%%%%%%%%%%%%%%%%%%%%%%%%%%%%%%%%%%%%%%%%%%%%%%%%

%%%%%%%%%%%%%%%%%%%%%%%%%%%%%%%%%%%%%%%%%%%%%%%%%%%%%%%%%%%%%%%%%%%%%%
% The bibliography is stored in an external database file
% in the BibTeX format (file_name.bib).  The bibliography is
% created by the following command and it will appear in this
% position in the document. You may, of course, create your
% own bibliography by using thebibliography environment as in
%
% \begin{thebibliography}{12}
% ...
% \bibitem{itemreference} D. E. Knudsen.
% {\em 1966 World Bnus Almanac.}
% {Permafrost Press, Novosibirsk.}
% ...
% \end{thebibliography}

% Here's where you specify the bibliography style file.
% The full file name for the bibliography style file 
% used for an ASME paper is asmems4.bst.
%\bibliographystyle{asmems4}

% Here's where you specify the bibliography database file.
% The full file name of the bibliography database for this
% article is asme2e.bib. The name for your database is up
% to you.
%\bibliography{asme2e}

%%%%%%%%%%%%%%%%%%%%%%%%%%%%%%%%%%%%%%%%%%%%%%%%%%%%%%%%%%%%%%%%%%%%%%

\end{document}